\documentclass[preprint,showpacs,preprintnumbers,amsmath,amssymb,nofootinbib]{revtex4}

\usepackage{amsthm}

{Corollary}[section]

\usepackage{etex}
\usepackage{amssymb,amsthm,amscd,amsbsy,array}
\usepackage{bm}
\usepackage{amsmath,dsfont}
\usepackage{soul} 
\usepackage{graphics,graphicx,xcolor}
\usepackage[colorlinks=true, pdfstartview=FitV, linkcolor=blue, citecolor=blue, urlcolor=blue]{hyperref} %
\usepackage{algorithm}
\usepackage{algorithmic}
\usepackage{longtable}
\usepackage{epstopdf}
\usepackage{subfigure}
\usepackage{float}
\usepackage[T1]{fontenc}
\usepackage[english]{babel}
\usepackage{graphicx}
\usepackage{float}
\usepackage{tikz}

\newcommand{\gb}{\colorbox{green}}

\newenvironment{redtext}{\color{red}}{\ignorespacesafterend}
\newenvironment{bluetext}{\color{blue}}{\ignorespacesafterend}

\newenvironment{magentatext}{\color{magenta}}{\ignorespacesafterend}
\newenvironment{orangetext}{\color{orange}}{\ignorespacesafterend}
\newenvironment{cyantext}{\color{cyan}}{\ignorespacesafterend}
\newcommand{\bblue}{\begin{bluetext}}
\newcommand{\eblue}{\end{bluetext}}
\newcommand{\bred}{\begin{redtext}}
\newcommand{\ered}{\end{redtext}}
\newcommand{\bmagenta}{\begin{magentatext}}
\newcommand{\emagenta}{\end{magentatext}}
\newcommand{\borange}{\begin{orangetext}}
\newcommand{\eorange}{\end{orangetext}}
\newcommand{\bcyan}{\begin{cyantext}}
\newcommand{\ecyan}{\end{cyantext}}

\numberwithin{equation}{section}

\let\ssection=\section
\renewcommand{\section}{\setcounter{equation}{0}\ssection}

\newcommand{\cA}{{\mathcal{A}}}

\newcommand{\PT}{{P\"oschl{\strut}-Teller\;}}
\newcommand{\GW}{{gravitational wave\;}}
\newcommand{\GWs}{{gravitational waves\;}}

\newcommand{\bX}{{\bf X}}

\def\smallover#1/#2{\hbox{$\textstyle\frac{#1}{#2}$}} %

\def\besub{\begin{subequations}}
\def\esub{\end{subequations}}
\def\benu{\begin{enumerate}}
\def\eenu{\end{enumerate}}
\def\beq{\begin{equation}}
\def\eeq{\end{equation}}
\def\beqa{\begin{eqnarray}}
\def\eeqa{\end{eqnarray}}

\def\barray{\left(\begin{array}}
\def\earray{\end{array}\right)}
\def\barraynb{\begin{array}}
\def\earraynb{\end{array}}

\def\?{\quad{\gb{\fbox{\texttt{?}}\;}}\quad}

\def\v0{\mathbf{0}}

\def\beq{\begin{equation}}
\def\eeq{\end{equation}}
\def\bea{\begin{eqnarray}}
\def\eea{\end{eqnarray}}

\def\6{\partial}
\def\7{\tilde}
\def\8{\widehat}

\def\G11{\Gamma_{11} }

\newcommand{\const}{\mathop{\rm const.}\nolimits}
\newcommand{\half }{\frac{1}{2}}

\def\smallover#1/#2{\hbox{$\textstyle\frac{#1}{#2}$}} %
\def\smallcirc{{\raise 0.5pt \hbox{$\scriptstyle\circ$}}}
\def\2{{\smallover1/2}}
\def\aand{{\quad\text{\small and}\quad}}

\def\ie{{\;\text{\small i.e.}\;}}
\def\ie,{{\;\text{\small i.e.,}\;}}

\newcommand{\fm}{\mathfrak{m}}

\let\ssection=\section
\renewcommand{\section}{\setcounter{equation}{0}\ssection}

\begin{document}


\title{
 Gravitational wave memory ~: further examples \footnote{
Talk given by P-M Zhang on July 15 at the {\sl 33rd/35th International Colloquium on Group Theoretical Methods in Physics} (ICGTMP, Group33/35),
held in Cotonou, Benin, July 15 - 19, (2024).
https://icgtmp.sciencesconf.org/ }}

\author{
P.-M. Zhang$^{1}$\footnote{mailto:zhangpm5@mail.sysu.edu.cn},
Q.-L. Zhao$^{1}$\footnote{mailto: zhaoqliang@mail2.sysu.edu.cn},
M. Elbistan${}^{2}$\footnote{mailto: elbistan@itu.edu.tr},
P. A. Horvathy$^{3}$\footnote{mailto:horvathy@univ-tours.fr}
}

\affiliation{
$^1$ School of Physics and Astronomy, Sun Yat-sen University, Zhuhai, China
\\
$^2$ Department of Energy Systems Engineering, Istanbul, Bilgi University, 
 Turkey
\\
${}^{3}$ Institut Denis-Poisson CNRS/UMR 7013 - Universit\'e de Tours - Universit\'e d'Orl\'eans Parc de Grammont, 37200; Tours, France 
\\
}
\date{\today}


\begin{abstract}
Ehlers and Kundt [1]
argued in favor of the velocity effect: particles initally at rest hit by a burst of gravitational waves should fly apart with constant velocity  after the wave has passed. Zel'dovich and Polnarev [2] suggested instead that waves generated by flyby would be merely displaced. Their prediction is confirmed provided the wave parameters take some particular values.
\bigskip

key words: 
Plane Gravitational waves; Memory Effect; Displacement effect
\end{abstract}

\maketitle

\tableofcontents

\section{Introduction}\label{Intro}

The Velocity Memory (VM) effect was first considered by Ehlers and Kundt \cite{Ehlers} who argued that particles initally at rest hit by a burst of gravitational waves would fly apart with constant velocity \cite{Ehlers}. Their investigations were continued 
 in \cite{GibbHaw71,BraTho,BraGri,ZelPol,Sou73,GriPol}. 
Zel'dovich and Polnarev \cite{ZelPol} suggested in particular that
 \GWs  generated by flyby would exhibit the {Displacement Effect} (DM)~: the particles would merely be displaced.  
 Our earlier investigations  
 \cite{ShortMemory,LongMemory,EZHRev} agreed with the general scheme 
 \cite{ZelPol,GibbHaw71,BraTho,BraGri,ZelPol,Sou73,GriPol}
 but did not confirm  their claim, as seen in 
 FIG. \# 12. of ref. \cite{EZHRev}.
 However, recent study \cite{DM-1, Jibril, DM-2} indicates that 
 \emph{for certain ``miraculous'' values of the wave parameters}  we  \emph{do get DM}. 
In this Note we illustrate the theory by several examples.

We start with the $4$ dimensional, linearly polarized Brinkmann metric  \cite{Brink,DBKP,DGH91},
\beq
g_{\mu\nu}dX^\mu dX^\nu=
\delta_{ij} dX^i dX^j + 2 dU dV + 
\half{\cA}(U)\Big((X^1)^2-(X^2)^2\Big)(dU)^2\,,
\label{Bmetric}
\eeq
where $\bX=(X^i)$ are transverse coordinates and $U,\, V$ are light-cone coordinates. Its
geodesics are given by,
\begin{subequations}
\begin{align}
&\dfrac {d^2\!X^1}{dU^2}-\frac{1}{2}\cA X^1 = 0\,,
\qquad 
\dfrac {d^2\! X^2}{dU^2} + \frac{1}{2}\cA X^2 = 0\,,
\label{geoX1X2}
\\[6pt]
&\dfrac {d^2\!V}{dU^2} +\frac{1}{4}\dfrac{d\cA}{dU}\Big((X^1)^2-(X^2)^2\Big) 
+ 
\cA\Big(X^1\frac{dX^1}{dU}-X^2\dfrac{dX^2}{dU}\Big)=0\,.
\label{geoV}
\end{align}
\label{Bgeoeqn}
\end{subequations}
\noindent 
Having solved first eqns \eqref{geoX1X2}, \eqref{geoV} is solved  by   lifting  to the \GW spacetime \cite{DM-1,DM-2}. 
Henceforth we focus our attention at \eqref{geoX1X2}.
We recall that geodesic motion posses a conserved quantity referred to as the Jacobi invariant \cite{Eisenhart,EDAHKZ},
\beq
{\fm}^2 = -g_{\mu\nu}\dot{X}^{\mu}\dot{X}^{\nu} = \const 
\label{Jacobiinv}
\eeq
Discarding tachions, we shall assume that ${\fm}^2 \leq0$.

\section{The Memory effect}\label{MemorySec}

\subsection{Gaussian or \PT profile}\label{GPTSec}

 The Memory effect for Gaussian and \PT \cite{PTeller} profiles,
 \begin{equation}
\cA \equiv \cA^G(U) = \frac{k}{\sqrt{\pi}}e^{-U^{2}}
\aand
\cA^{PT}(U) = \dfrac{k}{2\cosh^2 U}\,,
\label{A0GPT}
\end{equation}%
was studied in
\cite{DM-1,DM-2,Chakra}. DM can be achieved in the attractive, but not in the repulsive coordinate sector~: we get ``half DM''. 

\subsection{$|U|^{-4}$ profile}\label{U-4}

Another example is obtained by considering  the 
 Brinkmann metric  \eqref{Bmetric} with profile \cite{Andrzejewski:2018zby,Andrzejewski:2018pwq}, 
\beq
\cA(U) = \frac{1}{{U}^4}\,.
\label{LPP-pulse}
\eeq
This is an approximate sandwich wave  both for $U < 0$ and $U > 0$ but is singular at  $U=0$. 
The profile  is  even  w.r.t. sign reversal,
\beq
U \to -U\,.
\label{Ureversal}
\eeq 
The sector of $X^1$ is repulsive,  that of $X^2$ sector is attractive. 

The geodesics of \eqref{LPP-pulse} can be found analytically. We recall first the linearly polarized vacuum \GW of Brdicka \cite{Brdicka}, \eqref{Bmetric} with  profile $\cA(U)=\const$
 Its geodesics correspond to the Bargmann lift \cite{DBKP,DGH91,EZHRev,DM-1} of the transverse trajectories
\begin{eqnarray}
	\widetilde{X}^1(\widetilde{U})=c_1\cosh(\widetilde{U})+c_2\sinh(\widetilde{U}), \qquad \widetilde{X}^2(\widetilde{U})=c_3\cos(\widetilde{U})+c_4\sin(\widetilde{U})\,. 
	\label{sol-1}
\end{eqnarray}
Then the conformal redefinition  \cite{Gibbons2014,ZZH22}
\begin{eqnarray}
	U=-\frac{1}{\widetilde{U}}, \quad \bm{X}=\frac{\widetilde{\bm{X}}}{\widetilde{U}}, \quad
	V=\widetilde{V}+\frac{\widetilde{\bm{X}}^2}{2\widetilde{U}}\,, 
	\label{trans}
\end{eqnarray}
carries the Brdicka metric 
 conformally into \eqref{Bmetric} with profile \eqref{LPP-pulse}\,,
whereas  \eqref{sol-1} is taken into,
\begin{eqnarray}
	{X^1}({U})
	= &
	c_1\,{U}\cosh\left(\displaystyle\frac{1}{{U}}\right)
	-c_2\,{U}\sinh\left(\displaystyle\frac{1}{{U}}\right)
	\,,
	\label{rLPP-sol-X} 
	\\[6pt]
	{X^2}({U})
	= &
	c_3\,{U}\cos\left(\displaystyle\frac{1}{{U}}\right)
	- c_4\,{U}\sin\left(\displaystyle\frac{1}{{U}}\right)\quad
	\,.
	\label{rLPP-sol-Y}
\end{eqnarray}

\begin{figure}[htbp]
		\centering
	\subfigure[\label{even-X-U4}]{
		\begin{minipage}[t]{0.32\linewidth}
			\centering
			\includegraphics[scale=.32]{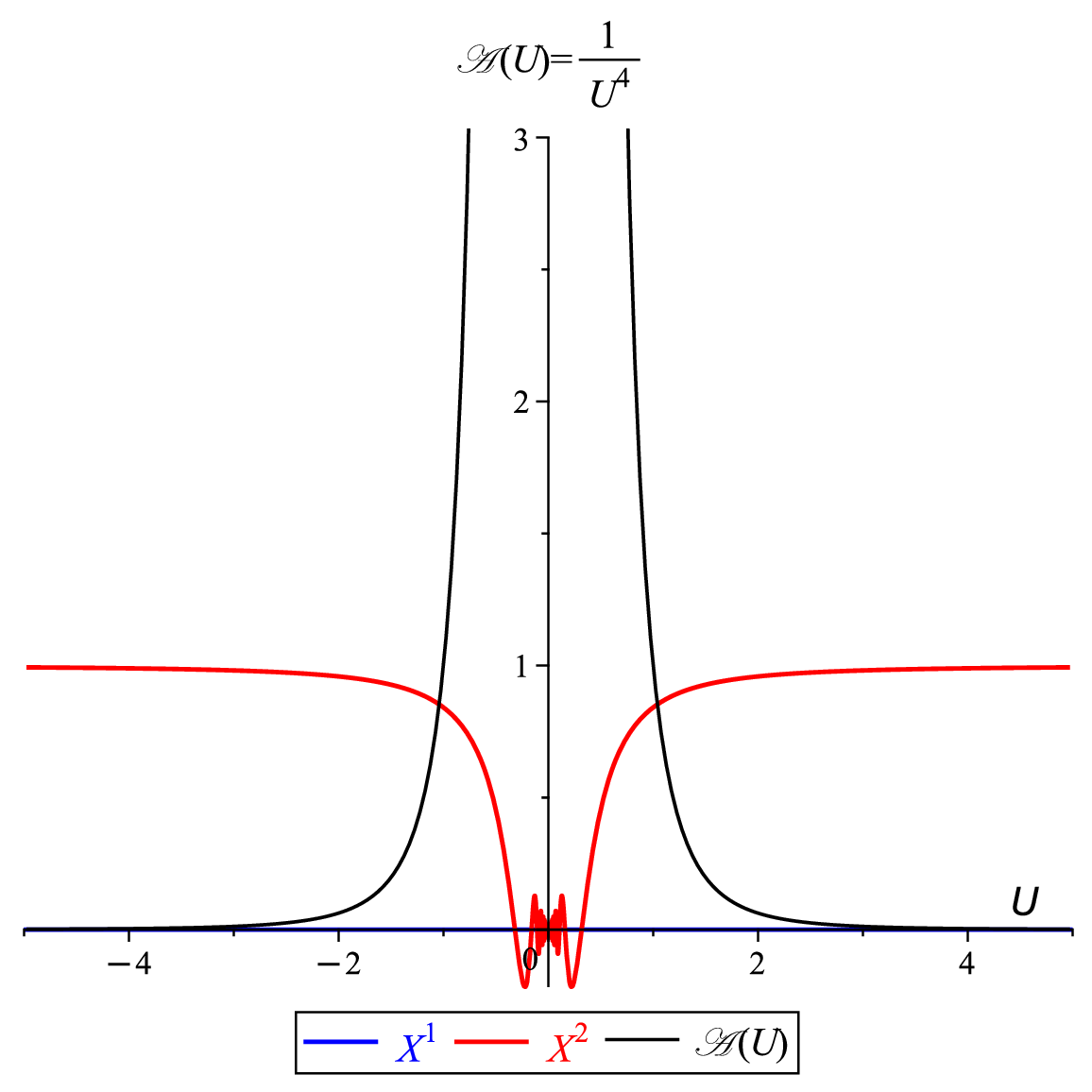}
		\end{minipage}
	}\qquad\qquad\qquad
	\subfigure[\label{even-V-U4}]{
		\begin{minipage}[t]{0.32\linewidth}
			\centering
			\includegraphics[scale=.32]{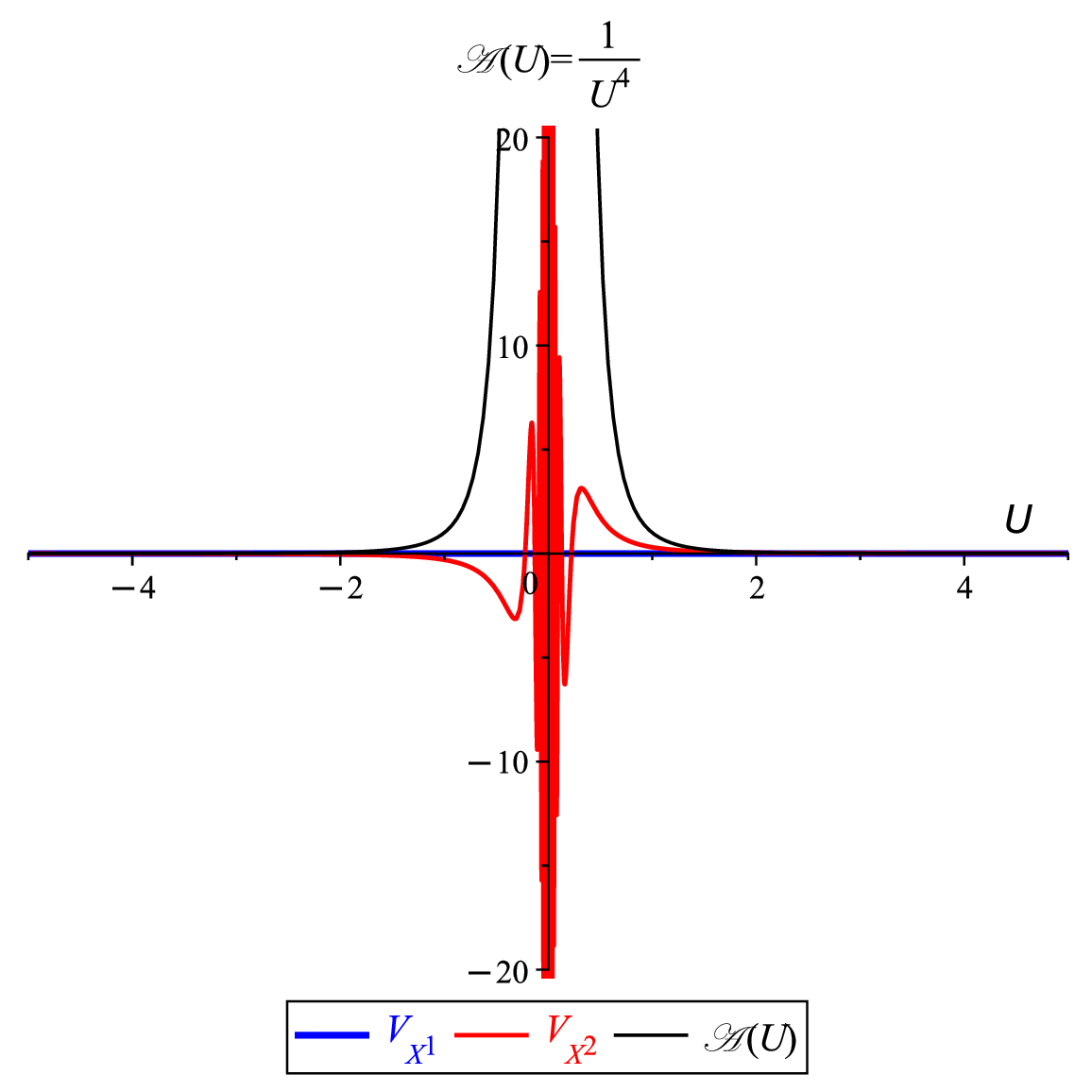}
		\end{minipage}
	}
	\caption{\textit{\small 
			\ref{even-X-U4}: For $c_2=1$ as only nonzero parameter, the evenly-glued $X^2$ component \eqref{evenglue} is symmetric w.r.t. $U$-reversal \eqref{Ureversal} and  thus  $X(-\infty) = 1 = X(+\infty)$. 
			\ref{even-V-U4}: 
			At $U=\pm\infty$ the velocity 
			goes to zero as it should.
		}
		\label{even-U4}
	}
\end{figure}

\begin{figure}[htbp]
	\centering
	\subfigure[\label{rLPP-displace-X-2-1}]{
		\begin{minipage}[t]{0.33\linewidth}
			\centering
			\includegraphics[scale=.33]{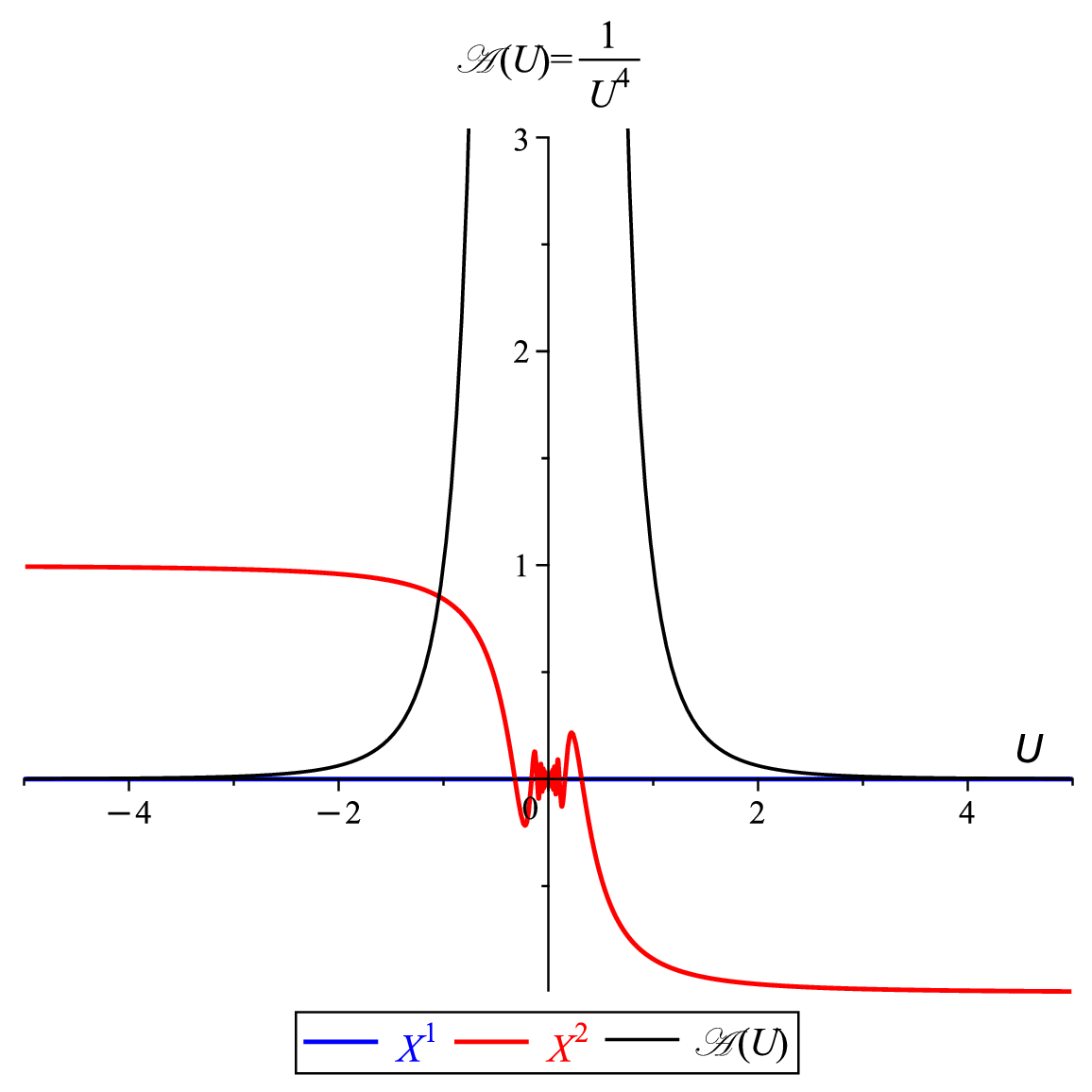}
		\end{minipage}
	}\qquad\qquad\qquad
	\subfigure[\label{rLPP-displace-VX-2-1}]{
		\begin{minipage}[t]{0.35\linewidth}
			\centering
			\includegraphics[scale=.35]{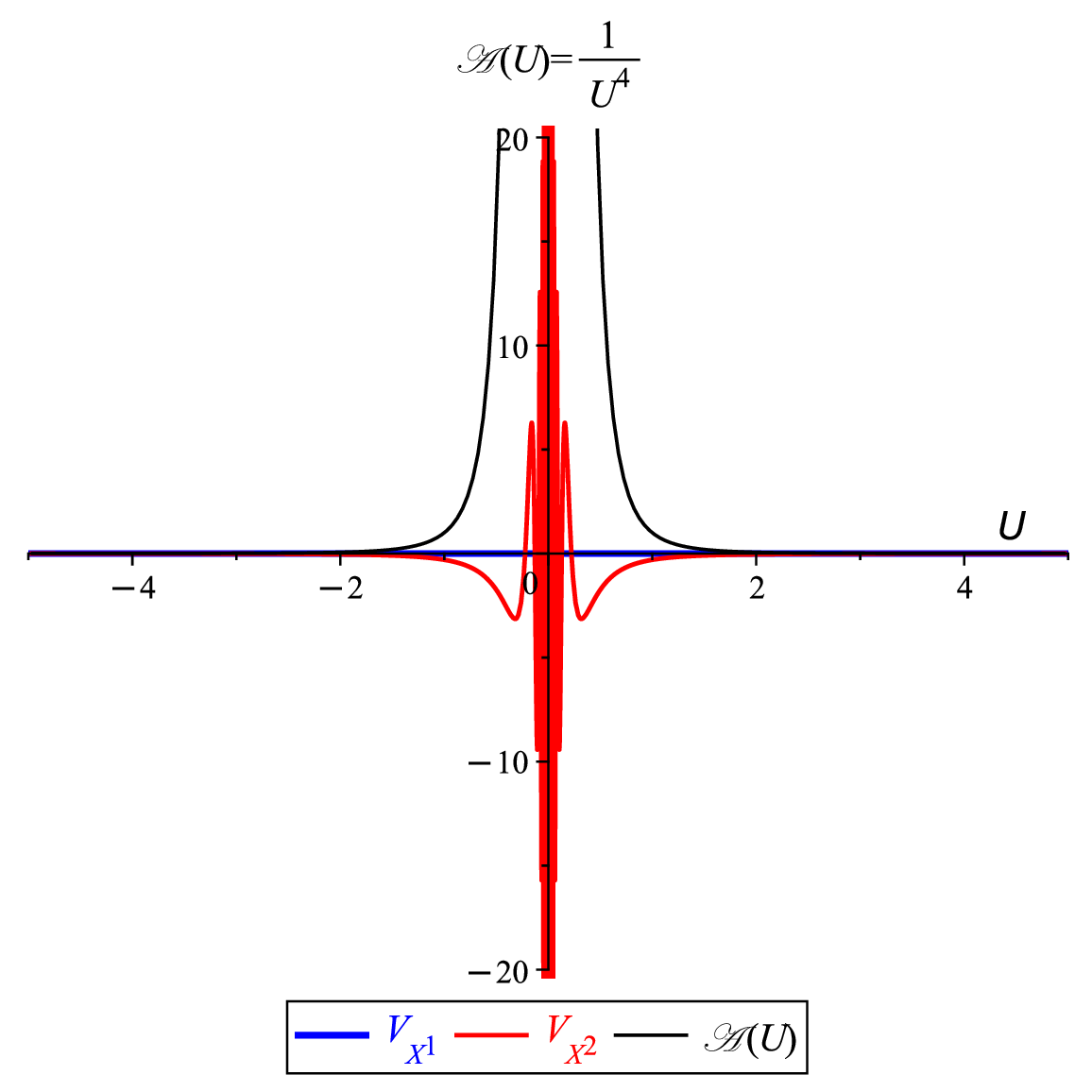}
		\end{minipage}
	}
	\\
	\caption{\textit{\small 
	\ref{rLPP-displace-X-2-1}: For $c_2=1$ as only nonzero parameter, the oddly-glued  component \eqref{oddglue} is antisymmetric w.r.t. $U$-reversal \eqref{Ureversal} and is thus displaced from $X^2(-\infty) = 1$ to $X^2(+\infty) =-1$. 
	\ref{rLPP-displace-VX-2-1}: 
	At $U=\pm\infty$ the velocity 
	 goes to zero, consistently with DM.
}
\label{rLPP-displace-X}
	}
\end{figure}

The terms with coefficients $c_1$, $c_3$ are odd and those with coefficients $c_2$, $c_4$ are even w.r.t. $U$-reversal, \eqref{Ureversal}.
 The singularity at $U=0$ can be removed when we posit
\beq
X^1(0) = X^2(0) = 0\,,
\label{0value}
\eeq 
allowing  us to glue together the $U<0$ and $U>0$ branches. 
However we have a curious  freedom for gluing,
  which can yield either even or odd trajectories. Let us first look at the attractive, $X^2$ sector. 
Starting with the even term for $U < 0$ and positing \eqref{0value}, we can continue by the \emph{same} expression also for $U > 0$, 
\begin{eqnarray}
	X^2_{even}({U})=\left\{
	\begin{aligned}
		&\, c_2\,{U}\sin\left(\frac{1}{U}\right) \quad U<0 
	\\
		& \quad  0 \quad U=0 
	\\
		&\, c_2\,{U}\sin\left(\frac{1}{U}\right) \quad U>0
	\end{aligned}
    \right.\; ,
    \label{evenglue}
\end{eqnarray}
and get an even trajectory, shown in FIG.\ref{even-U4}. 
However, starting with the same branch for $U \leq 0$ we can continue it for $U> 0$ also with \emph{minus} the same expression, \ie,
\begin{eqnarray}
	X^2_{odd}({U})=\left\{
	\begin{aligned}
		&\;\;\; c_2\,{U}\sin\left(\frac{1}{U}\right) \quad \;\; U<0 \\
		& \; \; \; \quad 0 \qquad \qquad \qquad  U=0 
	\\
		& -c_2\,{U}\sin\left(\frac{1}{U}\right) \quad U>0
	\end{aligned}
    \right.\quad ,
    \label{oddglue}
\end{eqnarray}
shown in FIG. \ref{rLPP-displace-X-2-1}. 
Both of these trajectories are continuous but non-differentiable at $U=0$, consistently with the velocity figures FIGs.\ref{even-V-U4} and \ref{rLPP-displace-VX-2-1}.

 \emph{Both} the evenly and the oddly - glued cases,
\eqref{evenglue} and \eqref{oddglue}, exhibit DM. 
However the  different ways of glueing result in that their outgoing values alternate,
\beq
\lim_{U\to \infty}X^2_{even}(U)= \; 1  
\aand
 \lim_{U\to \infty}X^2_{odd}(U)= -1 \,,
\label{e-o-DM}
\eeq
consistently with the behavior w.r.t. parity \cite{DM-2}.

\begin{figure}[htbp]
	\centering
	\subfigure[\label{rLPP-ana-sol-Y-1}]{
		\begin{minipage}[t]{0.3\linewidth}
			\centering
		\includegraphics[width=6cm]{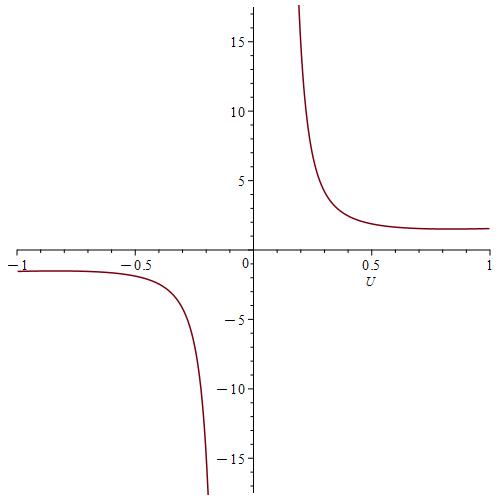}
		\end{minipage}
	}\qquad\qquad
	\subfigure[\label{rLPP-ana-sol-Y-2}]{
		\begin{minipage}[t]{0.3\linewidth}
			\centering
			\includegraphics[width=6cm]{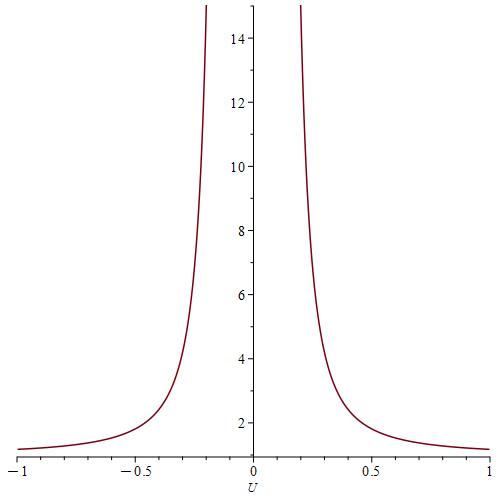}
		\end{minipage}
	}
	\\
	\caption{\textit{\small 
	Both of the $c_2=1$ odd \ref{rLPP-ana-sol-Y-1} 
	and the $c_1=1$ even solution \ref{rLPP-ana-sol-Y-2} diverge at $U=0$
	requiring $X^1(U)\equiv 0$ for all $U$.
	}
	\label{rLPP-ana-sol-Y}
	}
\end{figure}

Consistently with the general pattern in  \cite{DM-2}, no DM is possible in the repulsive, $X^1$ sector. 
Both the motion   \eqref{rLPP-sol-Y} and its velocity, shown in FIG.~\ref{rLPP-ana-sol-Y}, diverge at $U=0$ and can not be regularized unless $c_1=c_2=0$ i.e., when $X^1(U)\equiv0$~: we get ``half DM''.
\goodbreak

\subsection{Flyby (and beyond)}
\label{dGPT}
	
Flyby \cite{ZelPol}  was thoroughly analysed  for both the Gibbons - Hawking \cite{GibbHaw71} and  the \PT  \cite{PTeller} derivative-profiles \cite{DM-1,DM-2}, 
\begin{equation}
\cA^{G}=
\frac{\;\;\,d}{dU}\left(\frac{k}{\sqrt{\pi}}%
e^{-U^{2}}\right)\,
\aand
\cA^{PT}=
\frac{\;\;\,d}{dU}\left(
\frac{k}{2\cosh^2U}\right)\,,
\label{A1GPT}
\end{equation}%
respectively. Odd parity, $\cA^{G}(-U) = -\cA^{G}(-U)$, implies that it is enough to consider one component only. Then DM is obtained in \emph{both} coordinate sectors \cite{DM-2}.

\begin{figure}[h]
\includegraphics[scale=.6]{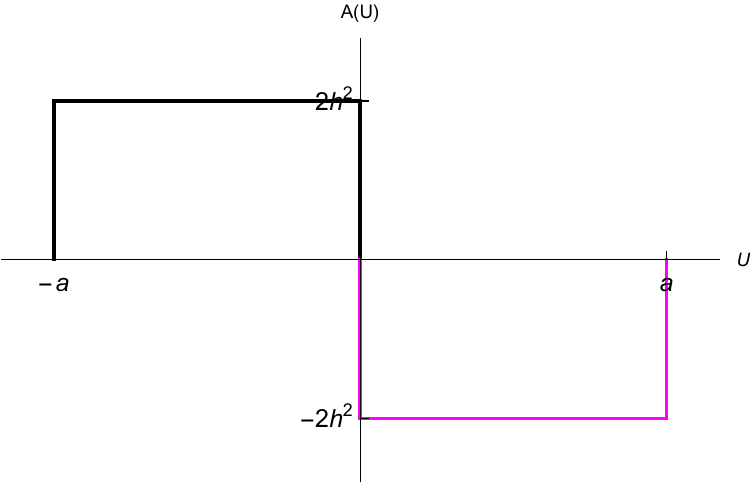} 
\vskip-4mm
\caption{\textit{Double-square approximation of the flyby profile.}
\label{fig1a}
}
\end{figure}

No analytic solutions are known for the  profile $\cA^{G}$ in \eqref{A1GPT} proposed by Gibbons and Hawking \cite{GibbHaw71}. 
For the derivative of the \PT profile, $\cA^{dPT}$ we do get
analytic solutions, however they involve highly sophisticated manipulations with  special functions  \cite{DM-2}. 
Our clue is that the profiles in \eqref{A1GPT} can, although rather crudely, be approximated by the double-square potential,
\begin{equation}
\cA\left( U\right) =\left\{ 
\begin{array}{c}
0 \text{ \ \ \ \ \ \ \ \ }U\ <-a \\ 
\;2h^{2} \text{ \ \ }-a<U\leq 0 \\ 
-2h^{2} \text{ \ \ \ }0<U<a \\ 
0,\text{ \ \ \ \ \ \ \ \ \ }U>a%
\end{array}%
\right. 
\end{equation}
 shown in FIG.\ref{fig1a}, which generalizes \cite{ChakraII}.
The initial conditions are choosen as,%
\begin{equation}
X\left( 
U \leq -a\right) = X_{0}\text{, \ \ \ \ }V_{0}=\frac{dX}{dU}\Big|_{U \leq  -a}=0\,.
\label{leftinit}
\end{equation}%
In the region $-a \leq U \leq0$ the profile is that of an inverted oscillator, 
$\ddot{X}-h^{2}X=0$. Thus, 
\begin{equation}
X_{I}\left(U\right) =X_{0}\cosh \left(hU+ha\right) .
\end{equation}

At the right boundary $U=0$ we have, 
\begin{equation}
X_{1}:=X_{I}(U=0)=X_{0}\cosh \left( ha\right),   
\quad
V_{1}:=\frac{dX_{I}}{dU}(U=0)=hX_{0}\sinh \left( ha\right)\,  
\label{xdx10}
\end{equation}
which serve as initial conditions in the right region $0\leq U \leq a$, where we have  a harmonic oscillator combining its  $\sin, \cos$ solutions.
 DM requires  the velocity to vanish for $U=a$. %
This can be achieved  exponentially, 
when 
\begin{equation}
ha=m\pi +\frac{\pi}{4}\,,\qquad m=0,1,2,3,\, \dots
\label{hampi}
\end{equation}%
Finally we  find DM asymptotically, %
\begin{equation}
X\left( U\right) =\left\{ 
\begin{array}{lllc}
X_{I}(U)=&X_{0}\cosh \left(hU+ha\right), 
&-a<U<0 
\\[6pt] 
X_{II}\left( U\right) =&X_{0}\left[ \cosh \left( ha\right) \cos \left(
hU\right) +\sinh \left( ha\right) \sin \left( hU\right) \right] ,\quad
&\quad 0<U<a%
\end{array}%
\right. 
\end{equation}%
This behavior is reminiscent of the one we found for derivative-\PT.

Higher-derivative profiles  can be also  considered \cite{GibbHaw71,LongMemory}. The 2nd derivative of the Gaussian (or of \PT\!)
describe, e.g. the  Braginsky - Thorne system  \cite{BraTho,GibbHaw71}, for which we get half DM.
The 3rd derivative was proposed instead
to describe \emph{gravitational collapse} \cite{GibbHaw71}, 
for which we get  again \emph{DM} for both components \cite{DM-2}. 

\section{Conclusion}\label{Concl}

 Zel'dovich and Polnarev suggested  that flyby would generate   pure displacement ({DM}) with vanishing relative velocity \cite{ZelPol}. Our papers \cite{DM-1,DM-2} indicate that for 
  \emph{judicious choices of the wave parameters} 
\emph{does} yield approximately pure displacement, refining the statement of Zel'dovich and Polnarev. 

At last, Sect. 2-5.8 of \cite{Ehlers} anticipates much of later work.  Similar problems were considered also in ref. \cite{AiBalasin,Mitman,HarteOancea,DeyKar}.
\goodbreak
 
\vskip-7mm
\begin{acknowledgments}\vskip-5mm
Some of our results were obtained in collaboration with J. Balog. Discussions are acknowledged to P.~C. Aichelburg, T. Damour, L. Di\'osi and  G.~Gibbons. 
 PAH thanks the Erwin Schr\"odinger Institute (ESI, Vienna) for hospitality during the Workshop
 {\sl Carrollian Physics and Holography}${}_{-}$CDFG${}_{-}$2024. PMZ was partially supported by the National Natural Science Foundation of China (Grant No. 12375084). M.E. is supported by Istanbul Bilgi University research fund BAP with grant no: 2024.01.00.

\end{acknowledgments}
\goodbreak



\end{document}